\documentstyle[aps,prl,psfig,twocolumn,floats]{revtex}
\begin{document}

\draft

\title{Non-Linear and Non-Local Meissner Effect\\ in Superconducting Wires}

\author{J. S\'anchez-Ca\~nizares,${}^1$ J. Ferrer,${}^2$ and F. Sols$\,{}^1$}

\address{
${}^1$Departamento de F\'{\i}sica Te\'orica de la Materia Condensada
and \\
Instituto de Ciencia de Materiales ``Nicol\'as Cabrera''\\
Universidad Aut\'onoma de Madrid, E-28049 Madrid, Spain\\
${}^2$ Departamento de F\'{\i}sica, Facultad de Ciencias, Universidad de
Oviedo,
E-33007 Oviedo, Spain}


\address{\begin{minipage}[t]{6.0in}
\begin{abstract}
The structure of the Meissner effect in a current-carrying
cylindrical wire with arbitrary disorder is studied following a numerical 
procedure that is exact within the quasiclassical approximation. A
distribution of current is found that is non-monotonous as a
function of the radial coordinate. For high currents, a robust
gapless superconducting state develops at the surface of both
clean and dirty leads. Our calculation provides a quantitative theory of
the critical
current in realistic wires.
\end{abstract}
\end{minipage}}

\maketitle
The Meissner effect is one of the most fundamental properties of
the superconducting state. It originates from the existence of a
disipationless current density generated by the superfluid
velocity distribution which characterizes the state of 
broken gauge symmetry. The current density ${\bf j}$ depends on the velocity 
${\bf v}_s$
through a complicated functional ${\bf j}[{\bf v}_s]$ which in
general is non-linear and non-local \cite{bardeen62}. The
Pippard approximation assumes a linear relation
${\bf j}({\bf r})= \int \,{\bf K}({\bf r},{\bf r'}) \cdot
{\bf v}_s ({\bf r'}) \, d{\bf r'}$, which is valid for low enough superfluid 
velocities \cite{degennes_book} and which in the local limit reduces to London's 
equation. A local but yet non-linear approximation 
to the full functional ${\bf j}[{\bf v}_s]$ is implemented, for
instance, within the context of a Ginzburg-Landau description
\cite{degennes_book}.

In this Letter, we present a fully non-linear and non-local
numerical study of the field and current distributions in a
current-carrying cylindrical wire. Our calculation of the full
functional ${\bf j}[{\bf v}_s]$ is exact within the framework of
the quasiclassical approximation. We encounter a rich physical
structure determined by the interplay between non-linearity,
non-locality, and the global stability of the current
configuration. In particular we find that, as a function of the
distance to the central axis, the current density is
non-monotonous for currents close to the critical value, displaying a 
maximum near the surface. This configuration can be viewed as 
precursor of the intermediate state \cite{degennes_book}.
We also find that, for high
total currents, the superfluid velocity near the surface acquires
values so large that they cannot be realized in a
quasi--one-dimensional wire. In a three-dimensional wire, these large
values of the superfluid velocity are possible because they are
supported by the global stability of the current distribution.
This causes a strong distortion of the local
quasiparticle spectrum, which then develops a robust gapless form. 
The calculation presented here provides a quantitative theory of the
critical current and represents an improvement over the
phenomenological Silsbee's criterion \cite{abrikosov_book}.

We investigate the structure of currents and fields in a
cylindrical wire made of a type I s-wave superconductor with an
arbitrary degree of disorder due to non-magnetic impurities. 
We have explored a broad range of temperatures and wire radii, and have
found that the most interesting physics appears at low temperatures
and for large wire radii (specifically, for radius 
$R > \xi_0, \lambda_0$, where $\xi_0$ and $\lambda_0$ are the zero 
temperature coherence and penetration lengths).
Thus we have focussed on wires that are wide
enough both to let the different physical magnitudes vary across
its section, and to prevent thermal or quantum phase slips from
taking place. 
We envisage a steady-state scenario where, in the
absence of externally applied fields, a disipationless current
distribution ${\bf j}[{\bf v}_s]({\bf r})$ flows through the wire 
generating a magnetic field ${\bf B}({\bf r})=(mc/e) 
{\bf \nabla}\times{\bf v}_s$, where the superfluid velocity is 
${\bf v}_s=(\hbar/2m)[{\bf \nabla} \phi -(2e/c\hbar){\bf A}]$. 
Amp\`ere's law then reads
\begin{equation}
\label{ampere}
\nabla^2 {\bf v}_s=\frac{4\pi e}{mc^2}\,{\bf j}[{\bf v}_s].
\end{equation}

At this point we make use of
the BCS theory of superconductivity
in its quasi-classical formulation 
\cite{eilenberger68,rammer88a,lambert98a,sauls99}.
Together with Eq. (\ref{ampere}), the following set of equations must be
solved self-consistently:
\begin{equation}
\label{motion}
\hbar\,{\bf v}_F \cdot {\bf \nabla_R} \check{g} =
i\,\left(\,E-\hbar {\bf k}_F\cdot{\bf v_s}\,\right) \,
[\,\check{\tau}_3,\check{g}\,]-[\,\check{\Sigma},\check{g}\,]
\end{equation}
\begin{equation}
\label{gap}
\Delta=-\frac{i\,U}{8} \int \frac{d{\bf \hat{p}}}{4\pi}\int_{-E_D}^{E_D}
dE \,{\rm Tr}[\,(\hat{\tau}_1-i\hat{\tau}_2)\,\hat{g}^K\,]
\end{equation}
\begin{equation}
\label{current}
{\bf j}=-\frac{e\,N_0\,v_F}{4}
\int dE \int \frac{d{\bf \hat{p}}}{4\pi} \,{\bf \hat{p}} \,
{\rm Tr}[\,\hat{\tau}_3\,\hat{g}^K\,].
\end{equation}
Here, $\check{g}({\bf r},{\bf \hat{p}},E)$ is the quasiclassical
fermion propagator in Keldysh space, whose components are $\hat{g}^R$,
$\hat{g}^A$, and $\hat{g}^K$ \cite{lambert98a}, 
$\check{\Sigma}=\check{\Sigma}_{\rm BCS}+
\check{\Sigma}_{\rm imp}$ is its self-energy, which has contributions 
\cite{rammer88a} from the pairing interaction  ($\check{\Sigma}_{\rm BCS}$)
and from impurities ($\check{\Sigma}_{\rm imp}$),
${\bf \hat{p}}$ is a unit vector in the direction of ${\bf k}_F$,
$\Delta$ is the modulus of the superconducting order parameter,
$U$ is the coupling constant, and $N_0$ is the normal density of states.
$\hat{\tau_i}$ are Pauli matrices and $\check{\tau}_3$ is a block-diagonal
matrix  with block entries like in $\hat{\tau}_3$.
Disorder is included through $\check{\Sigma}_{\rm imp}$ and
is characterized by the
dimensionless parameter $\Gamma=\hbar/\tau\Delta_0$, $\tau$ being
the elastic scattering time and $\Delta_0$  the zero temperature,
zero current superconducting gap. 
Equations (\ref{ampere}), (\ref{gap}) and (\ref{current}), 
together with the continuity equation,
can be derived as the time-independent equations satisfied by the extrema of 
the gauge-invariant 
action of an s-wave superconductor \cite{otterlo99a,ferrer99a} whose 
fermion propagator obeys Eq. (\ref{motion}).

We exploit the cylindrical symmetry of the problem and
consider configurations in which ${\bf j}$ points parallel to
the wire axis. The continuity equation can then be shown to
imply that ${\bf v}_s$ is also directed 
along the wire, so that the magnetic field ${\bf B}$ has only
angular component. Like the order parameter
$\Delta$, these three quantities depend only on the radial
coordinate $\rho$.


\begin{figure}[p]
\psfigurepath{figures}
\psfig{figure=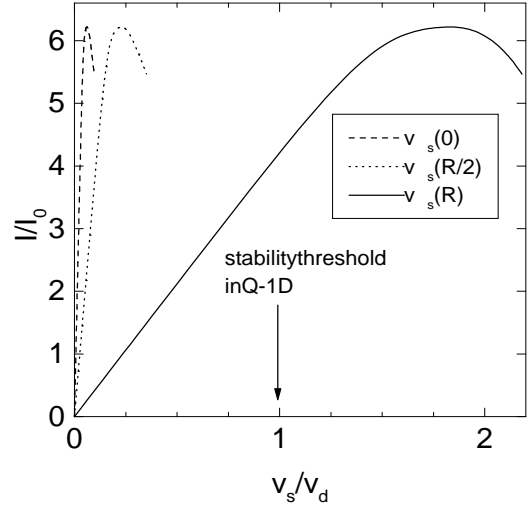,height=11cm,width=8cm,angle=0}
\vspace{-3.4cm} \caption{Total current $I$, in units of $I_0\equiv
3\pi e N_0v_F\Delta_0\lambda_0^2$, plotted as a function of the
superfluid velocity $v_s(\rho)$, in units of $v_d \equiv
\Delta_0/p_F$, at three different values of $\rho$, for a clean
($\Gamma=0$) wire of radius $R \!=\! 5\lambda_0$ at a temperature $T=0.2
\,T_c$. }
\end{figure}



\begin{figure}[p]
\psfigurepath{figures}
\psfig{figure=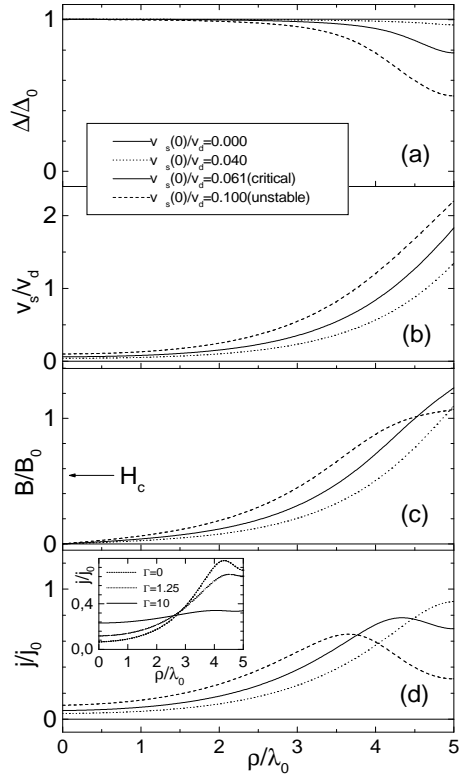,height=11cm,width=8cm,angle=0}
\caption{(a) Order parameter $\Delta$, (b) superfluid velocity
$v_s$, (c) magnetic field $B$, and (d) current density $j$, as a
function of the radial coordinate $\rho$, for the same wire as in
Fig. 1. The different current configurations are characterized by
$v_s(0)$. The units for current density and magnetic field are
$j_0 \equiv 2eN_0v_F \Delta_0 /3$ and $B_0 \equiv 6 \pi e N_0 v_F
\Delta_0 \lambda_0 /c$. The bulk critical field $H_c=0.54 B_0$ is
indicated in (c). Inset in (d): 
$j(\rho)$ at $I=I_c$ for several disorder strengths.}
\end{figure}


The self-consistency equations are solved according to the
following procedure: We assume initial profiles $\Delta(\rho)$ and
$v_s(\rho)$ for the order parameter and the superfluid velocity.
The quasiclassical fermion propagator is now obtained for all
values of $\rho,\hat{p},E$ by solving the equation of motion
(\ref{motion}) with hard-wall boundary conditions. This requires a 
self-consistent calculation of
$\check{\Sigma}_{\rm imp}(E)$ for all values of $E$. The resulting
$\hat{g}^K$ is introduced into Eqs. (\ref{gap}) and (\ref{current}) to
determine $\Delta(\rho)$ and $j(\rho)$. The
superfluid velocity $v_s(\rho)$ is then obtained by solving the 
differential equation (\ref{ampere}). The described calculation 
casts $\Delta(\rho)$ and $v_s(\rho)$ as output
distributions which are reintroduced as input for the next
iterative step. The whole procedure is repeated until
self-consistency in $\Delta$ and $v_s$ is achieved. The value of
$v_s(\rho=0)$ is given as an initial input parameter and is kept
fixed throughout the succesive iterative steps, effectively acting
as a label for the resulting current configuration.

Except where stated otherwise, we present
results for a superconducting
wire of radius $R=5\,\lambda_0$ at a temperature $T=0.2\,T_c$.
The Ginzburg-Landau parameter is $\kappa \equiv \lambda_0/\xi_0=0.4$ 
and the Debye energy is $E_D=10 \, \Delta_0$. 
In Fig. 1, we
plot the total current $I$ as a function of the superfluid
velocity $v_s$ at three different locations in the wire
($\rho=0,R/2$, and $R$). Current densities and velocities are
rather small in the core of the wire, even when $I$ equals its
critical value $I_c$. In contrast to this, $v_s$ can be quite
large near the surface. Fig. 1 indicates that it can amply exceed 
the stability threshold of a quasi--one-dimensional wire ($R \ll
\lambda_0,\xi_0$) made of the same material. These large values of
$v_s$ are possible because it is the global stability of the
current configuration what matters. Thus $v_s$ can be very high in
a small region of space, provided that it is sufficiently small in
other regions in such a way that the configuration is globally
stable. In Fig. 1 we plot $I(v_s)$ only for the clean ($\Gamma=0$)
case, but qualitatively similar results are obtained for dirty wires.

In Fig. 2 we present profiles of the order parameter,
superfluid velocity, magnetic field, and current density, for
several current configurations. For low currents, the order
parameter $\Delta(\rho)$ stays essentially flat, since $v_s$ is
small everywhere. As the current increases, $\Delta$ develops a
depression in the vicinity of the 
surface, brought about by the
large values which $v_s$ acquires in that region. Fig. 2(b) shows a
monotonous increase of $v_s$ as a function of $\rho$. This is
consistent with the constant sign of the magnetic field $B$ shown
in Fig. 2(c), since the two quantities are related by
$dv_s/d\rho=(e/mc)B$. In turn, $B(\rho)$ also shows a monotonous
increase which however tends to level off near the surface. 
Fig. 2(c) shows that, at the surface, $B$ can exceed the thermodynamic 
critical field $H_c$ by as much as a factor of 2, in clear violation of
Silsbee's criterion, which states that the critical current is
achieved when $B(\rho=R)$ equals $H_c$.


\begin{figure}[p]
\psfigurepath{figures}
\psfig{figure=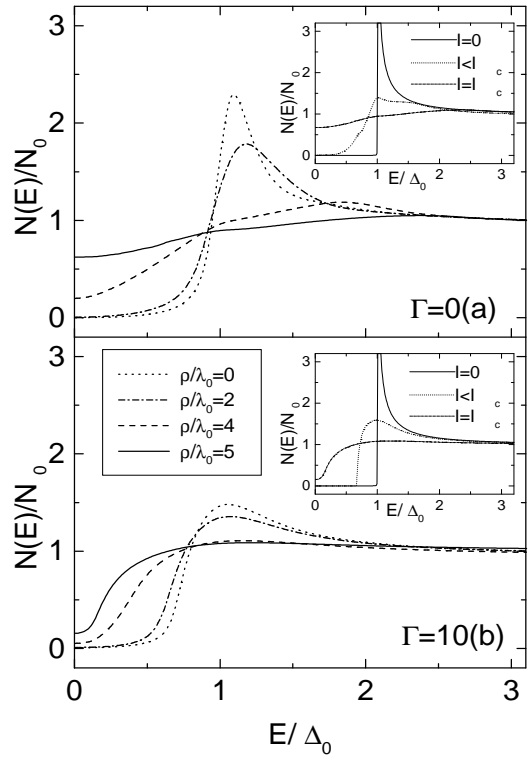,height=11cm,width=8cm,angle=0}
\vspace{0.0cm} \caption{Local density of states in the critical 
current configuration ($I_c$) at different distances from the 
central axis for (a) a clean and (b) a dirty wire. The insets show the LDOS 
at the surface of the wire in (a) the clean wire, for currents 
$I/I_0=0$, $2.90$, and $6.22$ (critical value), and in (b) the dirty 
wire, for currents  $I/I_0=0, 2.36$, and $3.92$ (critical value).}
\end{figure}



\begin{figure}[p]
\psfigurepath{figures}
\psfig{figure=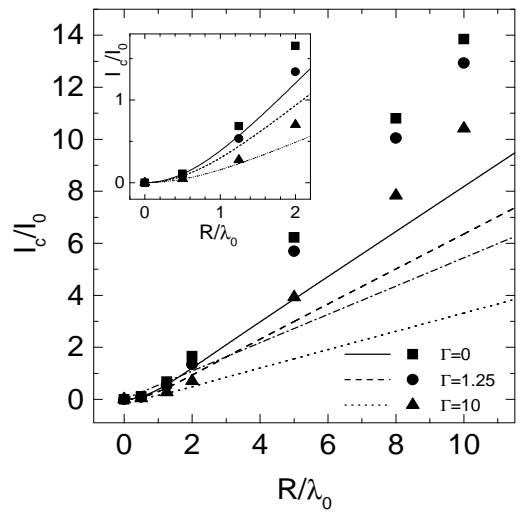,height=11cm,width=8cm,angle=0}
\vspace{-3.4cm} \caption{Critical current $I_c$ as a function of
radius $R$ for three different degrees of disorder. Discrete
points are the result of the exact calculation. Lines
are obtained from ansatz (\ref{ansatz}) in the main text
for several values of $\Gamma$. 
The dash-dotted line gives the prediction from Silsbee's rule. 
Inset: Magnification of the small $R$ quadratic behavior.}
\end{figure}

The large values which, for high current configurations, the
superfluid velocity is forced to adopt near the surface are
responsible for some of the most interesting physical features of
the Meissner problem. The current density displays a peculiar
behavior that can be approximately understood by noting that, in a
quasi--one-dimensional wire, $j$ is not a monotonous function of
$v_s$ but instead displays a maximum \cite{degennes_book,canizares99a}.
Thus, when $v_s$ becomes very large,
$j$ no longer increases with $v_s$ but actually decreases.
Translated to the three-dimensional wire, the consequence is that,
as a function of $\rho$, $j$ goes through a maximum 
near the surface. This can be clearly
appreciated in Fig. 2(d), whose inset shows in addition that a
similar (but yet less pronounced)
high-current behavior can be displayed in the presence of
appreciable disorder. We remark that this phenomenon is realized
for local values of $v_s$ that would be unstable (or directly not
realizable) in a quasi--one-dimensional wire.

The explanation given above suggests that the peak in the current
density profile can be understood in terms of a strictly local
theory which invokes the $j(v_s)$ relation of the quasi--one-dimensional wire.
This is the case, for instance, in a Ginzburg-Landau description, 
where the local relation $j(v_s) \propto
v_s(1-v_s^2)$ holds. We can check the adequacy of the local 
hypothesis, by considering the ansatz
$j \propto |\Delta|^2
v_s$, with $\Delta$ and $v_s$  taken from the exact calculation.
The result (not shown) is that, while the qualitatively correct 
behavior is reproduced in the dirty limit, the peak structure disappears in the 
clean case. The conclusion is that
an adequate description of the peak structure in the
current profile of a wire requires in general the combined inclusion of
non-locality and non-linearity in an essential way.

From the experimental viewpoint, the most relevant result reported
in this article is the peculiar behavior of the quasiparticle
density of states $N(E)$ at the surface of the wire. Close to the
surface, and for high total currents,
the local density of states (LDOS) strongly deviates from its 
conventional BCS shape, to the point of losing the coherence peak and becoming
gapless. This is shown in Fig. 3, where we plot the LDOS at
different positions in the wire for the critical current configuration
($I=I_c$), as well as for different currents right at the surface (see insets).
This strong distortion of the quasiparticle spectrum is caused by the large 
values which $v_s$ acquires near the surface when the total current is high
\cite{canizares99a,canizares97a}.
On the contrary, near the center of the wire, where $v_s$
is always small [see Fig. 2(b)], $N(E)$ approaches the standard BCS form.
In the dirty case, the fading of the gap at the surface is less pronounced, 
with the maximum value of $N(0)$ considerably smaller than in the clean case. 
This behavior occurs because disorder competes with
$v_s$ by tending to restore the gap in the LDOS
\cite{canizares99a,belzig98a}. As a result, $N(E)$ is less sensitive to the
$v_s$, hence showing a weaker dependence on $\rho$ and $I$. 
Unlike in the quasi--one-dimensional case \cite{canizares99a}, 
a stable form of transport-induced gapless superconductivity is 
induced near the surface because of global stability. This
strong transport dependence of $N(E)$ at the boundary could be
measured by performing a tunneling experiment on the surface of
current-carrying wire, in the spirit of experiments made on
superconducting films \cite{pyun89a}. We note that this is an
intrinsic surface effect that will survive for wires of
arbitrarily large radius.

Fig. 4 shows the critical current $I_c$ as a function of the radius $R$ of
the wire. A crossover from quadratic to linear behavior can be
clearly appreciated as the radius increases beyond the penetration
length. This crossover reflects the transition from
quasi--one-dimensional to full three-dimensional behavior.
Without the support of a fully self-consistent calculation like 
that presented here, an educated guess might be obtained from the assumptions
that $j$ depends linear and locally on $v_s$ and that the critical current 
$I_c(R)$ is reached when $j(\rho=R)=j_c(\Gamma)$, where 
$j_c(\Gamma)$ is the critical current density in a
quasi--one-dimensional wire with disorder $\Gamma$ \cite{canizares99a,kupryanov80a}.
Then we would obtain
\begin{equation}
\label{ansatz} I_c(R)= 2\pi \lambda_0 j_c(\Gamma)  R \,
\frac{{\rm I}_1(R/\lambda_0)}{{\rm I}_0(R/\lambda_0)}
,
\end{equation}
where $\rm{I}_0$ and $\rm{I}_1$ are modified Bessel functions of the
first kind. Fig. 4
shows that ansatz (\ref{ansatz}) works well for small $R$ (as
expected, since then it is exact), but systematically
underestimates $I_c$ for $R > \lambda_0$ in both clean and dirty
wires. Silsbee's rule also yields values for $I_c(R)$ which are 
too small \cite{comm1}.
The extra current which the superconductor accommodates as
compared with the predictions of more simple treatments, can be traced
to the non-monotonous radial dependence of the current density,
which allows for a bump near the surface. This result emphasizes
the need for a fully self-consistent calculation in the
formulation of a quantitative theory of the critical current.

In summary, we have calculated the structure of the intrinsic
Meissner effect in  realistic wires. Our description is based on a
numerical procedure which entirely incorporates the non-local and
non-linear aspects of the problem and which is exact within the
context of the quasiclassical formalism. The most relevant
features are the existence of a peak in the radial profile of the
current density near the surface and the generation of a
transport-induced gapless density of states at the surface, which could
be measured in a tunneling experiment. These
properties are displayed by stable high-current configurations in
both clean and dirty wires.
For wires of large radius, our calculation predicts values for the critical
current greater than those obtained from phenomenological models.


\acknowledgments

We thank W. Belzig, J. J. Palacios, and A. F. Volkov for valuable discussions.
This work has been supported by Direcci\'on General de Investigaci\'on
Cient\'{\i}fica y T\'ecnica under Grant No.
PB96-0080-C02, and by the EU TMR Programme under Contract No.
FMRX-CT96-0042.




\begin{thebibliography}{10}
\bibitem{comm1}
An improvement over this hypothesis within a Ginzburg-Landau context, 
for $\kappa \ll 1$, is given in Ref. \cite{abrikosov_book}.
\end{thebibliography}

\end{document}